\documentclass[runningheads]{llncs}
\usepackage{times}

\usepackage[T1]{fontenc}
\usepackage{graphicx}
\usepackage{color}
\usepackage{graphicx}
\usepackage{appendix}
\usepackage{multirow}
\usepackage{multicol}
\usepackage{hyperref}
\hypersetup{hidelinks, colorlinks=true, allcolors=blue, pdfstartview=Fit, breaklinks=true}
\usepackage{float}
\usepackage{comment}

\usepackage{breakcites}
\usepackage{xcolor}

\newcommand{\quo}[1]{``\emph{#1}''}

\usepackage{xspace}
\newcommand{\subsubtask}[1]{\vspace{0.5em}\noindent\textbf{#1}\xspace}

\begin{document}
\sloppy
\raggedbottom
%
\title{Turning Logs into Lumber: \\Preprocessing Tasks in Process Mining}
%
\titlerunning{Preprocessing Tasks in Process Mining}
%
\author{Ying Liu\inst{1}
\and
Vinicius Stein Dani\inst{1}
\and
Iris Beerepoot\inst{1}
\and
Xixi Lu \inst{1}
} 
%
%
\institute{Utrecht University, Utrecht, the Netherlands\\
\email{\{v.steindani, i.m.beerepoot, x.lu\}@uu.nl} 
}

\authorrunning{Liu et al.}
\maketitle              
\begin{abstract}
Event logs are invaluable for conducting process mining projects, offering insights into process improvement and data-driven decision-making. However, data quality issues affect the correctness and trustworthiness of these insights, making preprocessing tasks a necessity. 
%
Despite the recognized importance, the execution of preprocessing tasks remains ad-hoc, lacking support.
This paper presents a systematic literature review that establishes a comprehensive repository of preprocessing tasks and their usage in case studies. We identify six high-level and 20 low-level preprocessing tasks in case studies. Log filtering, transformation, and abstraction are commonly used, while log enriching, integration, and reduction are less frequent. These results can be considered a first step in contributing to more structured, transparent event log preprocessing, enhancing process mining reliability.
\end{abstract}


\section{Introduction}
\label{sec:intro}


%
%
In the landscape of data-driven decision-making, event logs stand as invaluable assets, capturing the execution of activities of processes and their interactions within diverse operational systems. The potential insights that can be obtained from these logs are immense, spanning process improvement, anomaly detection, performance evaluation, and strategic planning~\cite{Aalst2011}.
However, the axiom ``garbage in, garbage out'' holds particularly true in this context~\cite{DBLP:journals/is/SuriadiAHW17}. The presence of data quality issues underscores the vital importance of preprocessing techniques. Without proper preprocessing, the very foundation of analysis is compromised.

The importance of data quality and preprocessing in the field of process mining has been acknowledged, as evidenced by the growing attention dedicated to these subjects~\cite{DBLP:journals/is/SuriadiAHW17,van2021event}. Despite the acknowledgment, the execution of log preprocessing seems to remain ad-hoc. Moreover, little support has been provided on which preprocessing tasks are possible and how to select them. 
Although a few process mining methodologies sketched potential preprocessing tasks, a comprehensive overview of these tasks has been notably absent. Furthermore, the way these preprocessing tasks are used in real-life has remained unclear.

Existing systematic literature reviews (SLRs) have attempted to tackle specific tasks of log preprocessing, such as event abstraction techniques~\cite{van2021event} and data extraction~\cite{SteinDani2022Towards}. However, a comprehensive review that covers diverse preprocessing tasks and their practical applications in real-world scenarios is lacking.

In this paper, we perform a systematic literature review to establish an initial, comprehensive overview of the preprocessing tasks and their utilization in process mining case studies. By undertaking this endeavor, we aim to create a repository of log preprocessing tasks that may provide guidance and support for researchers and practitioners. 

We identified six high-level preprocessing tasks, and for four of these tasks, we observed 20 low-level preprocessing tasks described in the case studies. The results show that log filtering, transformation, and abstraction have been more frequently used in case studies, while log enriching, integration, and reduction (e.g., sampling) are much less frequently performed. 

The remainder of this paper is organized as follows. In Section~\ref{sec:relatedwork}, we discuss related work. Next, we explain the methodology followed in Section~\ref{sec:app} and present the results in Section~\ref{sec:results}. Finally, we conclude the paper in Section~\ref{sec:conclusion}.


\section{Related Work}
\label{sec:relatedwork}
\begin{figure}[h]
\makebox[\textwidth][c]{
    \includegraphics[width = \textwidth]{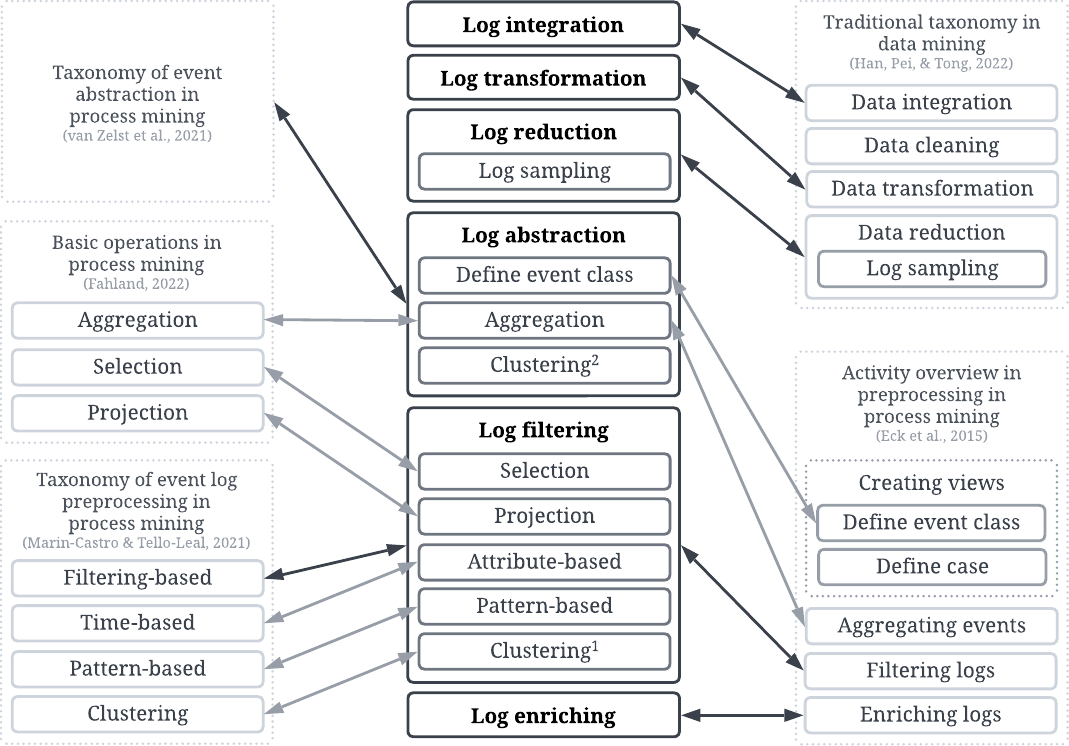}
}
    \caption{Initial result of high-level log preprocessing tasks and techniques in the related work.}
    \label{pre-results}
\end{figure}

In this section, we discuss the related work, based on which we synthesized an initial set of six high-level preprocessing tasks: (a) \emph{log integration}, (b) \emph{log transformation}, (c) \emph{log reduction}, (d) \emph{log abstraction}, (e) \emph{log filtering}, and (f) \emph{log enriching}, see Fig.~\ref{pre-results}.

\subsection{Taxonomy of Log Preprocessing Tasks}

Han et al. \cite{han2022data} propose four categories of data preprocessing techniques: data cleaning, data integration, data transformation, and data reduction. Data cleaning focuses on handling missing values, identifying noise or outliers, and repairing errors. Since these subtasks are not interesting (e.g., identifying missing variable values) or not directly applicable to process mining (e.g., identifying noise/outliers in a distribution), we omit this task and decide to focus on the latter three tasks. For each, we create a corresponding log preprocessing task: \emph{log integration}, \emph{log transformation}, and \emph{log reduction}.

Van Eck et al.~\cite{eck2015pm} listed four tasks in the preprocessing stage, which are specifically tailored towards event logs: creating views, aggregating events, filtering logs, and enriching logs. We exclude the task ``creating views'' because this task assumes that there is no event log yet, while we assume we have a raw event log as input. We match the task ``aggregating events'' to \emph{log abstraction}, also known as \emph{event abstraction} that has been already surveyed~\cite{van2021event}. The filtering of event logs (``filtering logs'') is also considered a log preprocessing task within our scope, which we refer to as \emph{log filtering}. Finally, the preprocessing task ``enriching logs'' is mapped to \emph{log enriching}.
As for \textit{log enriching} and \textit{log integration}, we consider \emph{log integration} as creating a new event log by integrating one or more external data sources, while \emph{log enriching} focuses on using the information within the event log to derive additional attributes.

Fahland~\cite{Fahland2022} indicated that there are three basic preprocessing operations on event logs, which are: selection, projection, and aggregation. We consider the ``selection'' and ``projection'' as a part of the \emph{log filtering} task, while the aggregation operation is considered as part of the \emph{log abstraction}. 

Regarding \emph{log filtering}, \emph{log abstraction}, and \emph{log reduction},
both \emph{log filtering} and \emph{log abstraction} can reduce the size of the logs, but we consider the following subtle differences in comparison to log reduction here. 
\textit{Log filtering} tends to focus on the quality issues of the original data. It obtains higher-quality logs by filtering out incorrect, incomplete, inconsistent, and irrelevant data. 
\textit{Log abstraction} focuses on the complexity and granularity of the original data. It groups the events through aggregation, defining event classes, and clustering to reduce the complexity of logs.
\textit{Log reduction} is due to the data volume of the original data. It reduces the amount of data processed in a single analysis by random sampling, dividing, or cutting, but still makes the data representative.

\subsection{Literature Review in Event Log Preprocessing} \label{LR in ELP}

To the best of our knowledge, there is only one literature review focusing on the log preprocessing tasks: Marin-Castro and Tello-Leal \cite{MarinCastro2021} reviewed 70 related papers that were published from the years 2005 to 2020 and explicitly mentioned event log preprocessing or cleaning. 
This literature review grouped preprocessing techniques into two types of \emph{techniques}: transformation techniques and detection-visualization techniques. \emph{Transformation techniques} mark modifications made toward the original structure of the event log, while the events or traces that can lead to issues with data quality are identified, grouped, and isolated using \emph{detection-visualization techniques}. In this paper, we cover six high-level \emph{preprocessing tasks}, instead of the techniques. We include log enriching, log integration, and log reduction, which have not been discussed. 

Van Zelst et al.~\cite{van2021event} conducted a review and presented a taxonomy of event abstraction techniques. While valuable and detailed insights are provided into the event abstraction techniques, no insights are provided into their usage in practice, and no overview is provided for other preprocessing tasks. 
%
%
Similarly, Stein Dani et al.~\cite{SteinDani2022Towards} report that preprocessing, on a high level represented by filtering-related tasks, is still a manual effort in the event log preparation phase of a process mining project. However, they mainly focus on data extraction tasks and do not provide an overview of the preprocessing tasks, including automated ones, and their usage in real-life case studies.

Currently, there is no clear overview of log preprocessing tasks and how frequently are these preprocessing tasks being used in process mining projects. Using the six high-level tasks as our scope, we conduct an SLR in order to provide insights into the usage of log preprocessing techniques in process mining case studies.



\section{Systematic Literature Review}
\label{sec:app}
\begin{figure}[h]
    \centering
    \includegraphics[clip, trim=0.2cm 5.2cm 0.2cm 0.23cm, width = \textwidth]{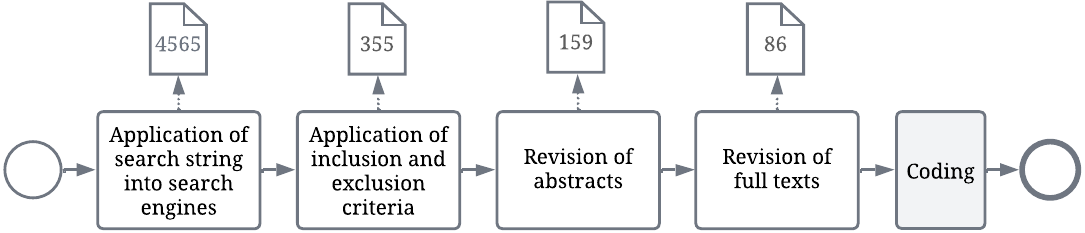}
    \caption{Paper screening procedure.}
    \label{paperscreen}
\end{figure}

To arrive at an initial selection of relevant papers, and inspired by Kitchenham and Petersen \cite{Kitchenham2009,Petersen2008SystematicMS}, we applied the following search string on Scopus: 
(``process mining'') AND (``case study'' OR ``case studies'') within the article title, abstract, and keywords. As of December 20, 2022, we initially found 4565 papers. Fig.~\ref{paperscreen} shows an overview of the paper screening process we followed. Next, we applied the exclusion and inclusion criteria in order to narrow down the scope of the review. 
The following exclusion criteria were defined and applied directly via the search engine: (1) the paper is published in 2021 and 2022; (2) the paper is published in conferences or journals under peer-review;~(3) the paper explicitly mentions ``process mining'' in the keywords; and, (4)~the paper is written in English.
As we are particularly interested in the current trend in case studies that use process mining as the core technique, this is our inclusion criteria. Therefore, only papers meeting these criteria were selected to be further analyzed.

After applying our exclusion and inclusion criteria, we obtained 355 papers. Because our focus is on log preprocessing tasks applied in real-world settings, we then read the abstracts of all these papers and filtered out the ones that did not mention collecting data from a real-world scenario. 
Thereafter, we obtained 159 papers to go through the full paper screening. These papers were downloaded and imported into the software Nvivo\footnote{https://lumivero.com/products/nvivo/} for further analysis. During the full paper analysis stage, the papers that did not mention any data preprocessing steps were discarded and, finally, 86 papers were obtained as relevant papers to go through the coding stage of our work.


The following codes were defined for the analysis: high-level category, low-level category, and data domain. Next, we discuss what each one of them entails.
\textit{High-level categories} were defined based on related work and used to deductively categorize the papers. The six high-level categories are (1) log integration, (2) log reduction, (3) log abstraction, (4) log filtering, (5) log enriching, and (6) log transformation.
%
Several \textit{Low-level categories} within the high-level categories were inductively defined from the studied papers.
%
Finally, in addition, we also coded the \textit{data domain} (e.g., healthcare, education, manufacture, etc.), the analysis purpose, the PM task, and the year. Due to space limits, we do not discuss these results in this paper. 
%
%
%
%
%
%
%
%
%
%
The initial result of the categorization process is presented in~Fig.~\ref{pre-results}.


\section{Results}
\label{sec:results}


\begin{table}[p]
    \centering
    \caption{Category citation details of 86 papers.}
    \label{citation detail} 
    \begin{tabular}{l| l |p{5cm} }
\hline
High-level category & Low-level category & References 
\\
\hline
 \multirow{11}{*}{Log filtering (55) }
 & Filtering irrelevant data (29) &
\cite{AcacioClaro2022,Ardimento2022,Bajo2022,Birk2021,Cenka2022,Chanifah2021,Chen2022a,Cuendet2022,Dupuis2022,goel2021improving,Grueger2022a,Hachicha2021,Ivanka2022,Jonk2022,Khaosanoi2021,LopezPernas2021,Pieters2022,Porouhan2022a,Rahmawati2022,Real2021,Rojas2021,Ruschel2021,SanchezSegura2022,Saralaya2022,stephan2021case,Tang2023,Tavazzi2021,theis2021improving,wisudiawan2022process}
\\
\cline{2-3}
~  & Filtering incomplete data (16) &
\cite{Bahaweres2021,cho2020discovery,Du2021,Dupuis2022,Gao2021,goel2021improving,Huda2021,Khaosanoi2021,Lamghari2022,Pang2021,Pieters2022,Pradana2022,rismanchian2023data,Rojas2021,Ruschel2021,Tavazzi2021}
\\
\cline{2-3}
~ & Filtering infrequent data (13) &
\cite{Bahaweres2022,Battineni2022,Chen2022,Dupuis2022,Hulzen2022,Ivanka2022,Kecht2023,Lamghari2022,Pradana2022,rismanchian2023data,Tariq2022,Tariq2022a,TavakoliZaniani2022}
\\
\cline{2-3}
~ & Filtering duplicates (8) &
\cite{Birk2021,Dogan2022a,Du2021,Dupuis2022,FahrenkrogPetersen2021,Pradana2022,Rojas2021,SanchezSegura2022}
\\
\cline{2-3}
~ & Filtering outliers (5) &
\cite{Benevento2022,Birk2021,Chen2022a,Lim2022,Ridwanah2022}
\\
\cline{2-3}
~ & Filtering incorrect data (4)&
\cite{Gao2021,Kropp2022,Pang2021, Song2022}
\\
\cline{2-3}
~ & Filtering redundant data (2) &
\cite{Chen2022,cho2020discovery}
\\
\cline{2-3}
~ & Filtering inconsistent data (1) & \cite{Chanifah2021}
\\
\cline{2-3}
~ & Filtering noise (3) &\cite{Chanifah2021,Leoni2022,RamosGutierrez2021}
\\
\hline
 \multirow{4}{*}{Log transformation (38) }
 & Transforming format (25) &
\cite{Ardimento2022,Battineni2020,beerepoot2021seeing,Birk2021,Cenka2022,Chanifah2021,DivyaSri2021,Du2021,Dupuis2022,Erdogan2022,Grueger2022,Hachicha2021,Husin2021,Ivanka2022,Khaosanoi2021,Kolakowska2022,martinez2021modelling,Pan2021,Pang2021,Pradana2022,Rahmawati2022,Ridwanah2022,Thiyagarajan2021,tridalestari2022analysis,Valensia2021}
\\
\cline{2-3}
~ & Transforming values (12) &
\cite{cho2020discovery,Dupuis2022,FahrenkrogPetersen2021,Kropp2022,Lamghari2022,Mehraby2022,Mertens2020,Pan2021,Pieters2022,Rojas2021,SanchezSegura2022,Saralaya2022}
\\
\cline{2-3}
~ & Reordering (5) &
\cite{Bajo2022,Du2021,Dupuis2022,LopezPernas2021,Pang2021}
\\
\cline{2-3}
~ & Transition matrices and encoding (2) & \cite{Dupuis2022,Yang2022}
\\
\hline
Log abstraction (37) & - & \cite{Adams2023,Ahmad2021,Battineni2022,Benevento2022,Cenka2022,Chen2022,Chen2022a,Cuendet2022,DivyaSri2021,Duma2023,Dupuis2022,Erdogan2022,Esposito2022,Grueger2022a,Husin2021,Koci2023,Kropp2021,Kropp2022,Lamghari2022,Leoni2022,LopezPernas2021,Macak2021,Macak2022,martinez2021modelling,Mehraby2022,Mertens2020,Pang2021,Porouhan2022a,Rashid2022,Revina2023,SanchezSegura2022,Tang2023,Tariq2022,Tariq2022a,Thiyagarajan2021,wisudiawan2022process,Yang2022}
\\
\hline
 \multirow{5}{*}{Log enriching (16)}
 & Adding calculation metrics (9) &
\cite{Dogan2022a,Duma2023,Huda2021,Kecht2023,Koci2023,Mehraby2022,Oberdorf2022,Revina2023,Schuh2021}
\\
\cline{2-3}
~ & Labelling (4) &
\cite{Araghi2022,Jonk2022,Pan2021,Tariq2022a}
\\
\cline{2-3}
~ & Adding case id (2) & 
\cite{Ridwanah2022,stephan2021case}
\\
\cline{2-3}
~ & Adding noise (1) &
\cite{Sohail2021}
\\
\hline
Log integration (14) &- &
\cite{Brockhoff2022,Cuendet2022,DivyaSri2021,Dogan2022a,Erdogan2022,goel2021improving,Hulzen2022,Kumbhar2022,Mertens2020,Oberdorf2022,Pradana2022,Ridwanah2022,SanchezSegura2022,Schuh2021}
\\
\hline
 \multirow{5}{*}{Log reduction (11)} 
 & Dividing into sub-logs (9) &
\cite{cho2020discovery,Esposito2022,goel2021improving,Huda2021,Kolakowska2022,Leoni2022,RamosGutierrez2021,Schuh2021,theis2021improving}
\\
\cline{2-3}
~ & Sampling (2) &
\cite{FahrenkrogPetersen2021,Song2022}
\\
\cline{2-3}
~ & Cutting traces (1) & \cite{FahrenkrogPetersen2021}
\\
\hline
\end{tabular}
\vspace{3em}
\end{table}

In this section, we present the results of coding 86 papers. The results are discussed for each high-level category. A complete overview of the results and the detailed coding can be found online, see the 
\href{https://docs.google.com/spreadsheets/d/1ScHe32-EFL7ZBR-7Vb4WiCFvlalQGIRgAfdAjzOgILM/edit?usp=sharing}{Google sheet file}. 
%
We also include the overview listed in Table~\ref{citation detail}.






\subsection{Log filtering} \label{filtering}

Log filtering is the most commonly performed preprocessing task, with 55 out of 86 papers performing this preprocessing task. These 55 papers mentioned filtering different objects, such as noise, outliers, redundant, duplicated cases and events, missing values, useless values, blank values, irrelevant values, and so on. Using the objects mentioned in these papers, the category log filtering is subdivided into 9 detailed low-level categories. 


\subsubtask{Filtering irrelevant data}
%
%
We observed that 29 out of 55 papers mentioned filtering ``irrelevant'' data. After analyzing these papers, we define \emph{irrelevant data} as \emph{those resources, activities, attributes, events, and traces that are not relevant or not important for the specific analysis to be conducted}.

Whether the data is relevant to the analysis task seems to be mostly determined by experts or analysts based on their domain knowledge and analysis requirements. For example, in \cite{Cenka2022}, the analysis only focused on the students who participated in the class (resource), so the events generated by other resources were defined as irrelevant data and filtered. In~\cite{goel2021improving}, the authors intended to analyze the activities of Ph.D. students and improve their journeys. So after a discussion by analysts and stakeholders, a filter is applied to retain the traces of full-time students who completed their Ph.D. and who withdrew (case status).
The term \emph{useless data} is also used in some of the papers to describe irrelevant data.
For example, in~\cite{Tang2023}, the authors mentioned \quo{filtering useless information such as links and marker symbols}, since the links and marker symbols (attributes) cannot make any contribution to the intended analysis and are regarded as useless data. 

\subsubtask{Filtering incomplete data}
In 16 out of 55 papers, the authors mentioned filtering incomplete data. Incomplete data can be divided into \emph{incomplete events} 
and \emph{incomplete cases}.
\emph{Incomplete events} usually refer to events having missing values or missing attributes. Incomplete events include missing case id \cite{Lamghari2022}, missing timestamps \cite{Du2021,goel2021improving,Pang2021}, and missing activities \cite{cho2020discovery,Du2021}, missing other attribute values that are relevant to this analysis \cite{Gao2021}.

The incompleteness of a case is usually described as cases that are not completed or do not represent the end-to-end process. It means that the cases lack some events, for example, \quo{remove any record that may create only one event per case as it will not depict the sequence of activities and hinder the performance analysis of the model}~\cite{Pradana2022} and \quo{removing cases that did not cover the whole steps}~\cite{cho2020discovery}. 

\subsubtask{Filtering infrequent data}
We use \emph{infrequent data} to refer to the infrequent case variant. In 13 papers, the authors mentioned that they performed the infrequent case variants filtering as a preprocessing task. Filtering infrequent data is done to \quo{prevent the PM tool from returning incomprehensible or inaccurate results}~\cite{rismanchian2023data}, and \quo{to improve the quality of results, and to avoid low precision and highly complex results}~\cite{TavakoliZaniani2022}.

\subsubtask{Filtering inconsistent data}
A simple example of inconsistent data is that the values are recorded in different formats, e.g., ``2023-01-01'' and ``2023/01/01'' as the attribute timestamps. This inconsistency in data format may be due to recording errors or caused by manual input. It may also be that different data sources have used different data formats. Inconsistent event labels make it difficult to assign clear semantics to the activities of a discovered process model~\cite{Aalst2011}, and may also bring about a dimensional explosion of the process model.


\subsubtask{Filtering incorrect data}
Incorrect data is erroneous or unreliable data that violates the logic of reality. For example, in the real process, activity $A$ should be executed earlier than activity $B$, but in the log, the timestamp of $A$ in a specific case is later than activity~$B$~\cite{Pang2021}.

\subsubtask{Filtering duplicates}
Duplicates refer to repeated data. In process mining, the case ID needs to be a unique identifier, and the traces represented by different case IDs must be different, so as to ensure the accuracy of the data. However, in real life, duplicate data is usually generated due to system bugs or other reasons. For example, in \cite{Dogan2022a}, repeated events with the same Call-ID were excluded.

\subsubtask{Filtering redundant data}
Only two papers mentioned redundant data \cite{Chen2022,cho2020discovery}. 
In \cite{cho2020discovery}, redundant events were included in data error: \quo{we conducted some data preprocessing, including handling data error (e.g., removing redundant events and eliminating multiple yield values)}, while there was no further definition and explanation in \cite{Chen2022}.

\subsubtask{Filtering outliers}
In \cite{Benevento2022,Birk2021,Lim2022}, the authors only mentioned \quo{removing outliers} without any further explanation or definition. In~\cite{Chen2022a}, the authors mention \quo{we noticed the existence of outliers, i.e., cases that take too long, or incomplete}; so, too long trace and incomplete data are considered outliers. 
In \cite{Ridwanah2022}, \quo{if lecture activities in the short semester are included, it will be an outlier because it has activities that are far more than short than activities in the semester in general}; thus, traces that are too short are also considered outliers. It seems that process analysts use the distributions of a case or event-attribute to define outliers, e.g., the number of events per case, the case duration per case, etc. 
%


\subsubtask{Filtering noise}
Noise is an overused word. Data that is not conducive to the analysis task is often defined as noise. An interesting point is that among the 86 papers, more than one paper mentioned noise, but only one paper described what noise is and how to filter it, \quo{In the original log the noisy activities were conveniently named `Noise', so they were removed using a filter on the activity name}~\cite{Leoni2022}.


\subsection{Log transformation}
In 38 of the 86 papers, the authors described that they performed a \emph{log transformation} task. The coding resulted in four data objects that are being transformed, which we use to further divide the high-level category.


\subsubtask{Transforming format}
Among the format transformations, the transformation of the log format from CSV to XES was mentioned the most (14 out of 25 papers), such that the event logs can be used in the PM tool. This is because the log format after extraction is usually CSV, and PM tools require the log format to be imported as XES.
The remaining format transformation is related to determining which columns are the key columns (such as case ID, activity name, and timestamps) after importing the log into PM tools.

\subsubtask{Transforming values}
The difference between transforming values and transforming format is that transforming values means the change of one or more specific values in an event. For example, replacing infrequent values with the value `other' to avoid dimension explosion, replacing missing values, replacing NaN values with `zero', capturing data, and encrypting data.

\subsubtask{Reordering}
Reordering is the process of sorting the log by a particular timestamp. When the original log is out of order, it is essential to reorder it so that the process model displays the activities' proper execution sequence.

\subsubtask{Transition matrices and encoding}
In particular, transition matrices and trace encoding are used as a preprocessing for predictive process monitoring. Given that the trace encoding is a subfield itself and was not included in the search, we consider this category outside of our scope. We found two case studies mentioning this preprocessing task and coded them without further analysis. 

\subsection{Log enriching}  \label{enriching}

\begin{figure}[tb]
\makebox[\textwidth][c]
{
    \includegraphics[width = 1\textwidth]{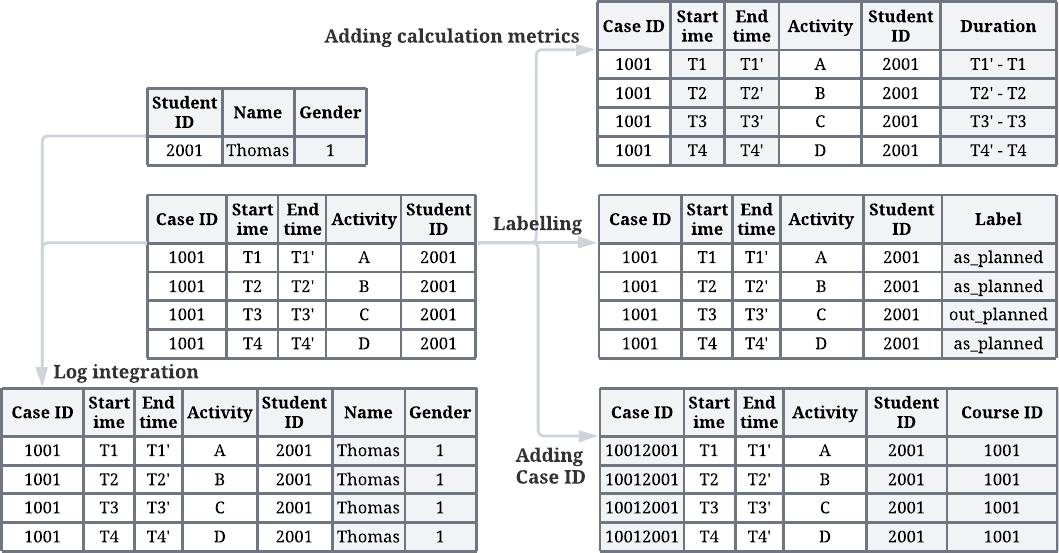}
}
    \caption{Examples of three log integration tasks versus log enriching.}
    \label{integration and enriching}
\end{figure}

In 16 out of 86 papers, the log-enriching techniques were applied. Log enriching is split into four categories. Three of them are shown using an example in Fig.~\ref{integration and enriching}.

\subsubtask{Adding calculation metrics}
In this low-level category, the calculation metrics are computed from existing attributes in the log. For example, in \cite{Dogan2022a}, call center processes of a company were examined. In the original event log, each call only had attributes Start and Call Duration, but process analysis required the end time of the call. Therefore, the attribute End was obtained by adding Call Duration to Start.



\subsubtask{Labelling}
Labeling is the task of assigning a tag or a class to an event or a trace. In~\cite{Tariq2022a}, \quo{the cases are labeled as either successful or failed, depending on how they have been executed and their outcome}, to further divide the log into two logs.
In~\cite{Pan2021}, for recording differences over time between the intended operation and the actual execution, a label was assigned to each event to indicate if the event was carried out on time or not.


\subsubtask{Adding case id}
Case id is a unique attribute in event logs. The data collected in some case studies did not have the attribute of case id, then the case id was created artificially in the data preprocessing stage. For example, in~\cite{stephan2021case}, ``the caseid is created by combining the three-digit client number (MANDT) with a ten-digit document number and a five-digit item number''.

\subsubtask{Adding noise}
Adding noise is not a typical preprocessing task, as just one publication described it.~\cite{Sohail2021} evaluated privacy assurance of healthcare metadata. Noise-adding plugins in the tool ProM were used to make the original event logs more privacy-preserving \cite{Mivule2013}.

\subsection{Log reduction}

\begin{figure}[tb]
\centering
    \includegraphics[width = \textwidth]{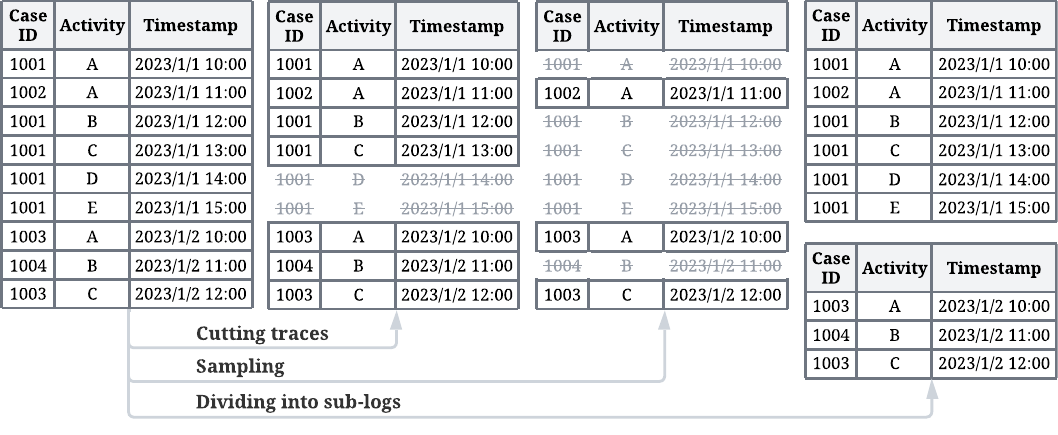}
    \caption{A simple example of log reduction.}
    \label{example-reduction}
\end{figure}

In 11 out of 86 papers, the authors used log reduction to do log preprocessing. 
Examples of the three log reduction tasks are shown in Fig.~\ref{example-reduction}.


\subsubtask{Dividing into sub-logs}
In the example presented in Fig.~\ref{example-reduction}, the original log is divided into two logs by the date in timestamp. In~\cite{Leoni2022,Esposito2022}, IoT logs were collected in a smart house and the aim was to explore human habits. They firstly divided logs into smaller pieces by timestamps to analyse the time distribution of the activities (user habits) within a day~\cite{Leoni2022}. 

Resource could also be a common attribute for division. The authors of~\cite{RamosGutierrez2021} divided the traces into subsets to model different profiles of users. Dividing original logs according to specific attributes is usually for more in-depth analysis~\cite{goel2021improving}.

In addition, in order to test the proposed algorithm or approach, the log was divided into training data and test data according to a certain proportion~\cite{cho2020discovery,Huda2021}.

\subsubtask{Sampling}
The most notable characteristic of sampling is randomness. The reduction here is to reduce the trace; that is to say, the data processing needs to be in the unit of a trace. In the example shown in Fig.~\ref{example-reduction}, there are four traces $[\langle A, B, C, D, E\rangle , \langle A\rangle,\langle A,C\rangle,\langle B\rangle]$. After randomly sampling 50\% of the traces, the log $[\langle A\rangle,\langle A,C\rangle ]$ in the lower right corner is obtained.

\subsubtask{Cutting traces}
In the example in Fig.~\ref{example-reduction}, compared to other traces, the trace $\langle A, B, C, D, E\rangle $ is obviously longer and contains more events. Cutting off the event at the end of the trace will get the processed log in the lower left corner. The purpose of this technique is to avoid bias from very long traces~\cite{FahrenkrogPetersen2021}.

\subsection{Log integration} \label{integration}
Among the 86 papers, 14 papers used log integration to combine multiple data tables. No objects of interest are repetitively mentioned, nor have we observed obvious low-level tasks. Therefore, the log integration task has not been further divided.  

Fig.~\ref{integration and enriching} shows an example where a new event log is created by matching two data tables using the shared attribute \quo{student\_id}. 
It is worth mentioning that some papers mention that additional data was added to the original event data without indicating the source, but we believe that the combination of these data is realized by log integration. According to~\cite{Hulzen2022}, \quo{Besides the attributes shown in Table 4, we included the educational level of the nurses executing the activity, as well as their nursing experience/organisational role, the hospital shift and weekday on which the activities were performed, and the ward in which the shift took place}. It is reasonable to speculate that this additional information actually comes from a separate data table that stores information about all nurses.

\subsection{Log abstraction}
In 37 out of the 86 papers, the authors used preprocessing techniques in log abstraction, which is the most widely performed task after log filtering and log transformation among the six preprocessing tasks. %
In~\cite{van2021event}, a review and taxonomy of event abstraction were presented. Therefore, we will not focus on this category here.  



\subsection{Discussion}

    %

    %
    %

%

    

The \emph{log filtering} task emerges as the most commonly performed preprocessing task, with over $63\%$ of the case studies mentioning that some filtering is performed. However, it's worth noting that the specifics of the log filtering tasks appear to heavily rely on domain knowledge. Moreover, more than 30 papers use somewhat ambiguous terminologies such as `irrelevant' or `noise'. 
The \emph{log transformation} task ranks as the second most frequently employed, accounting for $44\%$. Currently, the majority of subtasks in the log transformation focus on fixing format-related and data-quality issues. This highlights the importance of data quality in process mining and suggests that efforts to enhance data quality should continue to be a focal point in log preprocessing.

In contrast, log enriching~($18\%$), log integration~($16\%$), and log reduction~($12\%$) tasks are notably less commonly performed. One plausible explanation is the limited support for these tasks in both academic and commercial tools. Furthermore, the relatively uncommon use of log reduction can be attributed to the fact that many filtering techniques inherently reduce the log size. 








\section{Conclusion}
\label{sec:conclusion}
In this paper, we conducted a systematic literature review, examining the use of log preprocessing tasks in process mining case studies and presented the results. 
We identified six high-level tasks that were synthesized from the related work discussion and~20 low-level tasks inducted from the reported case studies. 
The log filtering task emerges as the most frequently used preprocessing task, featured in over~63\% of the case studies reviewed. The log transformation task follows closely behind, accounting for~44\% of the cases. Conversely, log enriching, integration, and reduction tasks are less commonly performed, possibly due to limited tool support. Future research can delve into these preprocessing tasks, providing operational guidance. Standardization in reporting practices and greater support for less common preprocessing tasks are valuable for improving traceability and advancing the reliability of process mining results.

 \bibliographystyle{splncs04}
 \bibliography{Thesis_bib}






\end{document}